\documentclass[12pt,american,preprint, aps,nofootinbib]{revtex4-1}
\usepackage[T1]{fontenc}
\usepackage[utf8]{inputenc}
\usepackage{babel}
\usepackage{mathrsfs}
\usepackage{amsmath}
\usepackage{amssymb}
\usepackage[pdfusetitle,
 bookmarks=true,bookmarksnumbered=false,bookmarksopen=false,
 breaklinks=false,pdfborder={0 0 1},backref=false,colorlinks=false]
 {hyperref}

\makeatletter
\usepackage{amsmath,amssymb,mathtools}
\newcommand{\AdS}{\mathrm{AdS}}
\newcommand{\dd}{\mathrm{d}}

\newcommand{\cA}{\mathcal{A}}
\newcommand{\Lie}{\mathcal{L}}

\makeatother

\begin{document}
\title{Gravitational Reconstruction in $\AdS_{4}$}
\author{David A. Lowe}
\author{Yiru Wang}
\author{Juanyi Yang}
\affiliation{Department of Physics, Brown University, Providence, RI 02912, USA}
\begin{abstract}
We reconstruct linearized gravitational perturbations in $\AdS_{4}$
from a conserved, traceless stress tensor on a fixed conformally flat
boundary metric, extending our earlier real Lorentzian construction.
For admissible data with no spacelike Fourier support in a flat conformal
frame, the physical response reduces to an integral over boundary
light-cone cuts. Boundary improvements restore Fefferman-{}-Graham
(FG) gauge while preserving normalizability and the linearized Weyl
tensor. We then reconstruct perturbations with general spacelike boundary
momentum, including nonzero frequency and its zero-frequency limit,
using explicit position-space generalized functions on real stress
data. Boundary kernels are constructed for normalizable perturbations
and sufficient conditions are derived for the reconstruction to converge.
The results motivate controlled flat-space limits relevant to Bondi-{}-Metzner-{}-Sachs
symmetries and celestial holography.
\end{abstract}
\maketitle

\section{Introduction\label{sec:intro}}

In the present work we reconsider the problem of reconstructing bulk
gravitational perturbations from boundary data in the context of Anti-de
Sitter/Conformal Field Theory duality (AdS/CFT) \citep{Maldacena:1999aa,Witten:1998qj}.
It follows from this duality conjecture that quantum gravity phenomena
in the bulk spacetime (AdS) may be mapped to well-defined quantum
field theory calculations in the CFT. This motivates understanding
in detail and simplifying this reconstruction procedure to as large
an extent as possible.

Here we extend the real Lorentzian reconstruction of Lowe, Wang and
Yang \citep{Lowe:2024reconstruction}. In that construction, conservation
and tracelessness reduce a boundary integral over a spacelike region
to its light-cone boundary, using stress data at real points. Here
we determine its normalizable source domain, extend reconstruction
to general spacelike boundary momentum, and derive the improvements
required for Fefferman-{}-Graham (FG) gauge in curved conformal representatives.
In both the patch and global constructions, these improvements leave
the linearized Weyl tensor unchanged. The extension to conformally
flat boundary metrics is local; the explicit global construction uses
the Einstein cylinder.

\subsection{Relation to earlier reconstruction formulas\label{sec:priorwork}}

HKLL reconstruction provides the broader framework \citep{Hamilton:2005ju,Hamilton_2006,Hamilton:2006fh,Hamilton:2007wj,Kabat:2011rz}.
Kabat et al.'s graviton formula, Eq. (37) of \citep{kabat}, integrates
$T_{ab}(t+t',\boldsymbol{x}+i\boldsymbol{y}')$ over $t'^{2}+|\boldsymbol{y}'|^{2}<z^{2}$,
using complexified spatial boundary coordinates. The real light-cone
integral of \citep{Lowe:2024reconstruction} instead acts on a time-dependent
source class defined by smoothness, spectral support, and endpoint
conditions, without assuming holomorphic continuation. The spin-two
radial equation and compensating diffeomorphisms remain common ingredients.
Section \ref{sec:spacelikecomplex} compares the radial profiles for
general spacelike momentum and distinguishes the real-data source
domain from the stronger assumptions needed for literal complex-coordinate
smearing.

Bhattacharjee and Mukherjee \citep{BhattacharjeeMukherjee:2025} derive
real Poincare kernels supported at spacelike-separated boundary points
for both graviton boundary branches by mode sums and Green's theorem.
Their mode expansion uses real $q=\sqrt{\omega^{2}-|\boldsymbol{k}|^{2}}$.
Here we retain the light-cone localization of \citep{Lowe:2024reconstruction}
and determine the boundary terms needed for its FG completion.

\subsection{Extensions of the real light-cone construction\label{sec:extensions}}

In the present work, we examine conditions on the boundary sources
which ensure the boundary integrals collapse to the light-cone-cut
form. Then we further correct the boundary kernel to ensure the resulting
bulk perturbation remains in FG form. Assume the boundary stress is
smooth and that its components and all derivatives fall off faster
than any inverse power at large boundary coordinates. In a flat conformal
frame, the Lorentzian kernel then selects a fixed boundary metric
precisely when the Fourier transform vanishes at spacelike momentum
(i.e., its support is restricted to timelike and null momenta). The
same restriction removes the lower endpoint in the FG improvement,
allowing the Ward identities to place both terms on the light-cone
cut.

Global normal modes require a separate prescription. The smooth-source
example in Appendix \ref{app:globalexamples} shows that an ordinary
distribution confined to the complete light-cone cuts cannot recover
the FG metric for every smooth source; the missing contribution is
a pure-gauge improvement. For time-analytic superpositions of the
regular global modes satisfying the bounds below, the full response
admits a convergent contact-term representation. The analytic single-mode
example in the same appendix also rules out unmodified antipodal gluing
of the patch weight.

The complementary open spacelike sector $|\boldsymbol{k}|^{2}>\omega^{2}$
includes both nonzero frequency and $\omega=0$ . Its standard normalizable
Bessel profile \citep{kabat,BhattacharjeeMukherjee:2025} grows exponentially
with spacelike momentum. We define a real position-space generalized
kernel on sources with quantitative exponential Fourier bounds, or
sufficient real derivative bounds, and derive its curvature-dependent
FG improvement. A covariant series of contact terms gives a local
FG kernel in any conformally flat boundary representative, provided
the transformed stress satisfies the derivative bounds and the FG
coordinates remain regular. Boundary normalizability is imposed independently
of interior regularity.

The boundary metric and stress are standard FG data \citep{deHaro:2000vlm},
and analytic-data convergence is established; see \citep{SerantesWithers:2022}
and references therein. Our results specify the real kernels and their
domains, including the global gauge obstruction and its resolution
for time-analytic sources. Fixing the normalizable branch is essential:
conservation and tracelessness alone do not impose this constraint,
and a boundary-preserving diffeomorphism cannot supply that condition.

The earlier real construction also relates boundary stresses to superrotation
charges \citep{Lowe:2020qan,Lowe:2024reconstruction}. CFT stress-tensor
Ward identities on the Lorentzian cylinder yield conformally soft
graviton relations \citep{deGioiaRaclariu:2024}, while a Carrollian
stress tensor encodes leading and subleading soft theorems at null
infinity \citep{BagchiDhivakarDutta:2026}. The general spacelike
family includes a controlled zero-frequency limit. Section \ref{sec:celestial}
explains how that limit and the regular radiative modes could be used
to test these connections through a scaled flat-space limit, with
the asymptotic data and charges kept explicit.

We work on exact $\AdS_{4}$, using locally conformally flat representatives
of its conformal boundary. Local results apply to any such representative;
the explicit global mode construction uses the Einstein cylinder.
Section \ref{sec:conventions} fixes conventions. Sections \ref{sec:lightcone}
-{}- \ref{sec:globaldynamic} develop the light-cone construction
and its global completion. Sections \ref{sec:spacelikeflat} and \ref{sec:spacelikeglobal}
give the spacelike extension, including its zero-frequency limit and
covariant FG kernel.

\section{Boundary data and conventions\label{sec:conventions}}

We set the AdS radius to one and use Lorentzian signature. The unperturbed
metric obeys $R_{\mu\nu}=-3G_{\mu\nu}$. We write $h_{\mu\nu}=\delta G_{\mu\nu}$
for the full metric perturbation and $z^{2}h_{ab}$ for its compactified
tangential perturbation. In FG coordinates,

\begin{equation}
G=z^{-2}\bigl(\dd z^{2}+\gamma_{ab}(z,y)\dd y^{a}\dd y^{b}\bigr),\qquad h_{zz}=h_{za}=0,\label{eq:FG}
\end{equation}
and the linearized compactified tangential metric has an expansion
\begin{equation}
z^{2}h_{ab}=A_{ab}+z^{2}A^{(2)}_{ab}+z^{3}B_{ab}+O(z^{4}).
\end{equation}

Fixing the boundary metric sets $A=0$ and its locally determined
companion $A^{(2)}=0$. The leading perturbation is then $h_{ab}=zB_{ab}+o(z)$.
We normalize the prescribed stress by

\begin{equation}
\tau_{ab}:=\frac{16\pi G_{4}}{3}\,\delta\langle T_{ab}\rangle=B_{ab},\qquad D^{a}\tau_{ab}=0,\qquad g^{ab}_{(0)}\tau_{ab}=0.\label{eq:tau}
\end{equation}

This is the standard odd-dimensional holographic normalization \citep{deHaro:2000vlm}.
A kernel acting on the physical stress therefore acquires a factor
$16\pi G_{4}/3$. If the source is realized as a thin layer at a radial
regulator, its strength must be scaled so that the boundary limit
gives the prescribed $B$. We refer to delta functions and their derivatives
as contact terms. We always specify their support, distinguishing
terms at coincident boundary points from terms on a boundary light-cone
cut.

We use the curvature convention in which the unit two-sphere has positive
Ricci tensor. For a locally conformally flat three-dimensional representative,
whose Cotton tensor vanishes, define its Schouten tensor and a radial
endomorphism by

\begin{equation}
P_{ab}=R_{ab}[g_{(0)}]-\frac{1}{4}R[g_{(0)}]g_{(0)ab},\qquad E_{a}{}^{b}(z)=\delta_{a}{}^{b}-\frac{z^{2}}{2}P_{a}{}^{b}.\label{eq:E}
\end{equation}

The exact AdS metric in FG coordinates is \citep{SkenderisSolodukhin:1999,deHaro:2000vlm}

\[
\gamma_{ab}(z)=E_{a}{}^{c}(z)E_{b}{}^{d}(z)g_{(0)cd}.
\]

This is a local description of exact AdS, valid where the boundary-normal
FG coordinate map is regular. Under $g_{(0)}=e^{2\omega}\chi^{*}\eta$
, the normalized covariant stress transforms as

\begin{equation}
\tau_{ab}=e^{-\omega}(\chi^{*}\widehat{\tau})_{ab},\qquad A_{ab}=e^{2\omega}(\chi^{*}\widehat{A})_{ab}.\label{eq:Weyl}
\end{equation}

Boundary dimension three has no local Weyl anomaly. Hats denote flat-frame
quantities. The condition $A=0$ is invariant under \eqref{eq:Weyl};
its Fourier characterization below refers to the chosen flat frame.

\subsection{Physical response and FG improvement terms\label{sec:physical}}

An improvement is a bulk diffeomorphism that puts a chosen normalizable
solution in FG gauge. For a pre-FG gauge perturbation $h^{\mathrm{pre\text{-}FG}}$,
write

\begin{equation}
h^{\mathrm{FG}}=h^{\mathrm{pre\text{-}FG}}+\mathcal{I}[h^{\mathrm{pre\text{-}FG}}],\qquad\mathcal{I}[h]=\Lie_{\xi}G.
\end{equation}

We choose $\xi$ to preserve the boundary representative and stress.
Since the AdS background has $C[G]=0$, its linearized Weyl tensor
$\mathcal{W}[h]:=\delta C[h]$ obeys

\begin{equation}
\mathcal{W}[h+\Lie_{\xi}G]=\mathcal{W}[h]+\Lie_{\xi}C[G]=\mathcal{W}[h].\label{eq:Weylinvariance}
\end{equation}

A local change of boundary representative is implemented by a bulk
coordinate transformation $\Psi$ . If $H$ is the flat-frame perturbation,
its physical curvature transforms directly, $\mathcal{W}[\Psi^{*}H]=\Psi^{*}\mathcal{W}[H]$
. The pulled-back metric generally acquires radial components \citep{kabat}
; the boundary-anchored improvement restores FG gauge and leaves that
curvature unchanged. Both curvature and metric can therefore be reconstructed
locally in a conformally flat boundary representative, with the metric
requiring the FG improvement and the appropriate source bounds. The
global Weyl tensor can likewise be calculated from the full pre-FG
gauge perturbation, including its radial components, or from gauge-invariant
master fields \citep{KodamaIshibashi:2003}. The FG improvement is
unnecessary for that calculation.

For comparison, the physical part of the D'Hoker et al. propagator
has the form

\begin{equation}
\begin{aligned}\mathcal{P}_{\mu\nu;\mu'\nu'}(X,X')={} & \bigl(\nabla_{\mu}\nabla_{\mu'}\mathcal{U}\,\nabla_{\nu}\nabla_{\nu'}\mathcal{U}+\nabla_{\mu}\nabla_{\nu'}\mathcal{U}\,\nabla_{\nu}\nabla_{\mu'}\mathcal{U}\bigr)\mathsf{G}(\mathcal{U})\\
 & +G_{\mu\nu}(X)G_{\mu'\nu'}(X')\mathsf{H}(\mathcal{U}).
\end{aligned}
\label{eq:physicalpropagator}
\end{equation}

Here $\mathcal{U}=\mathcal{U}(X,X')$ is the dimensionless chordal
invariant of the two bulk points, denoted by $u$ in that reference,
and $\mathsf{G},\mathsf{H}$ are the coefficients in their Eq. (3.9).
In the following section,$u$ denotes the Poincare radial coordinate.
Tracelessness removes the second structure. The additional gradient
terms in their full propagator generate observation-side diffeomorphisms
or reduce, by source conservation, to boundary terms \citep{DHoker:1999bve}.
These boundary terms must be checked when specifying the Lorentzian
reconstruction.

\section{Real light-cone reconstruction in a flat conformal frame\label{sec:lightcone}}

\subsection{A simple real boundary kernel\label{sec:flatpreFG}}

In a flat conformal frame, the Lorentzian representative of \citep{Lowe:2024reconstruction}
already satisfies FG gauge. Use Poincare coordinates $(u,x^{a})$,
where $u>0$ is the radial coordinate and the conformal boundary lies
at $u=0$. The background metric is $G=u^{-2}(\dd u^{2}+\eta_{ab}\dd x^{a}\dd x^{b})$,
with $\eta_{ab}=\operatorname{diag}(-1,1,1)$. Set $w^{a}=x^{a}-x'^{a}$,
and $Q_{u}=u^{2}+\eta_{ab}w^{a}w^{b}$. Write $c=3/(4\pi)$ and define

\begin{equation}
\begin{aligned}I^{(0)}_{ab}(u,x) & =\int_{\mathbb{R}^{1,2}}\dd^{3}x'\,\Theta(Q_{u})\widehat{\tau}_{ab}(x'),\\
I^{(1)}_{ab}(u,x) & =\int_{\mathbb{R}^{1,2}}\dd^{3}x'\,Q_{u}\Theta(Q_{u})\widehat{\tau}_{ab}(x').
\end{aligned}
\label{eq:moments}
\end{equation}
The flat tangential pre-FG gauge perturbation is

\begin{equation}
H_{ab}(u,x)=-\frac{c}{u^{2}}I^{(0)}_{ab}(u,x),\qquad H_{uu}=H_{ua}=0.\label{eq:flatH}
\end{equation}
On the source class below, \eqref{eq:flatH} solves the vacuum linearized
Einstein equation with timelike normalization $B=\widehat{\tau}$.
The domain $Q_{u}>0$ is the boundary region spacelike separated from
the observation point. Its boundary $Q_{u}=0$ consists of the future
and past light-cone cuts.

For this section, assume each stress component is smooth and that
it and every derivative falls off faster than any inverse power of
$\sqrt{t^{2}+x^{2}+y^{2}}$ as this quantity tends to infinity. We
call this rapid decay. These assumptions make the integrals and their
boundary limits well defined. Directly from \eqref{eq:flatH}, the
boundary-metric coefficient is

\begin{equation}
\widehat{A}_{ab}(x)=-cI^{(0)}_{ab}(0,x).\label{eq:Aendpoint}
\end{equation}
Hence a fixed boundary metric requires the spacelike-region moment
at $u=0$ to vanish, in addition to the Ward identities.

\subsection{Fourier characterization of normalizability\label{sec:spectral}}

Our Fourier convention is $\widetilde{\tau}(\omega,\boldsymbol{k})=\int\dd t\,\dd^{2}x\,e^{i\omega t-i\boldsymbol{k}\cdot\boldsymbol{x}}\tau(t,\boldsymbol{x})$,
with inverse measure $\dd\omega\dd^{2}k/(2\pi)^{3}$. Let $p^{2}=|\boldsymbol{k}|^{2}-\omega^{2}$.
Define the Fourier transform by inserting an exponential damping factor
and removing it after integration. Away from the null cone this gives

\begin{equation}
\mathcal{F}\{\Theta(\eta_{ab}w^{a}w^{b})\}(p)=\begin{cases}
-4\pi(p^{2})^{-3/2}, & p^{2}>0,\\
0, & p^{2}<0.
\end{cases}\label{eq:FTtheta}
\end{equation}

Possible contact terms on the null momentum cone must be treated before
applying this formula to singular Fourier data. They cause no ambiguity
for the smooth, rapidly decaying transforms considered here: if the
transform vanishes on the open spacelike cone, it and all its derivatives
also vanish on the null cone, including the origin. Within this class,

\begin{equation}
\widehat{A}=0\quad\Longleftrightarrow\quad\widetilde{\widehat{\tau}}_{ab}(\omega,\boldsymbol{k})=0\quad\hbox{for}|\boldsymbol{k}|^{2}>\omega^{2}.\label{eq:restriction}
\end{equation}

The spacelike multiplier in \eqref{eq:Aendpoint} is $3(p^{2})^{-3/2}$,
which is nonzero on that open cone. An example is a smooth Fourier
transform with compact support strictly inside the timelike cone,
satisfying the transverse and traceless algebraic Ward identities.
Its inverse transform and all derivatives have rapid decay, so the
endpoint terms vanish. The light-cone derivations apply to this example
and, more generally, to rapidly decaying stresses obeying \eqref{eq:restriction}.

In the Fourier formulas below, we write $z=u$ for the flat FG radial
coordinate. For a timelike mode set $s=\sqrt{\omega^{2}-|\boldsymbol{k}|^{2}}$.
The pre-FG gauge perturbation then gives

\begin{equation}
H_{ab}(z,p)=\frac{3}{s^{3}z^{2}}\bigl[\sin(sz)-sz\cos(sz)\bigr]\widehat{\tau}_{ab}(p)=\left[z-\frac{s^{2}z^{3}}{10}+O(z^{5})\right]\widehat{\tau}_{ab}(p).\label{eq:timelike}
\end{equation}
This gives a fixed boundary metric and the prescribed normalizable
coefficient. For $q=\sqrt{p^{2}}>0$, however, the same kernel gives

\begin{equation}
H_{ab}(z,p)=\frac{3(1+qz)e^{-qz}}{q^{3}z^{2}}\widehat{\tau}_{ab}(p),\qquad\widehat{A}_{ab}(p)=\frac{3}{q^{3}}\widehat{\tau}_{ab}(p).\label{eq:spacelike}
\end{equation}
The spacelike branch decays in the interior but changes the boundary
metric.

A nonzero smooth stress with compact support in boundary spacetime
cannot have its Fourier transform vanish on an open spacelike set:
the transform is analytic at all complex momenta, so vanishing on
such a set would force it to vanish everywhere. Compact support is
therefore not a substitute for \eqref{eq:restriction}. Sources concentrated
on the null momentum cone, time-periodic sources, and sources with
slower decay require a separate endpoint and convergence analysis.
The equivalence \eqref{eq:restriction} is proved here for smooth
stresses with the rapid decay just specified.

\subsection{Reduction to the two light-cone cuts\label{sec:Wardcuts}}

Integrating by parts against a conserved, traceless source gives two
identities, proved in Appendix \ref{app:FGcompletion} :

\begin{equation}
\begin{aligned}I^{(0)}_{ab}(u) & =-2\int\dd^{3}x'\,\delta(Q_{u})w_{(a}w^{c}\widehat{\tau}_{b)c},\\
I^{(1)}_{ab}(u) & =2\int\dd^{3}x'\,\delta(Q_{u})w_{a}w_{b}w^{c}w^{d}\widehat{\tau}_{cd}.
\end{aligned}
\label{eq:Wardcuts}
\end{equation}

Parentheses symmetrize with weight one-half. The factors $\delta(Q_{u})$
are contact terms on the two light-cone cuts. Both sheets $w^{0}=\pm\sqrt{u^{2}+|\boldsymbol{w}|^{2}}$
enter with positive Jacobians. The first identity localizes $H$ to
the light-cone cuts; the second will localize the weighted radial
integral entering its FG improvement.

The relevant lower endpoint is

\begin{equation}
C_{ab}(x):=I^{(1)}_{ab}(0,x),\qquad\Box C_{ab}=6I^{(0)}_{ab}(0,x).\label{eq:Cendpoint}
\end{equation}

Away from the null cone, the Fourier transform of $w^{2}\Theta(w^{2})$
is $24\pi(p^{2})^{-5/2}$ for spacelike momentum and zero for timelike
momentum. Thus $C_{ab}=0$ for the rapidly decaying sources obeying
\eqref{eq:restriction} . On a larger space, imposing only $I^{(0)}(0)=0$
makes $C$ a homogeneous boundary wave by \eqref{eq:Cendpoint} ;
it need not vanish. Including both time sheets does not by itself
cancel this endpoint. As an example, the conserved, traceless endpoint
example in Appendix \ref{app:FGcompletion} has weighted fluxes that
add.

\section{FG improvement terms in conformal charts\label{sec:FGcut}}

\subsection{A boundary-anchored improvement\label{sec:FGprojection}}

Locally, a bulk coordinate transformation associated with \eqref{eq:Weyl}
carries the flat physical solution to the chosen boundary representative.
Its linearized Weyl tensor transforms tensorially. The transformed
metric can acquire radial components, which we remove by the improvement
$\mathcal{I}=\Lie_{\xi}G$ . For a normalizable pre-FG gauge perturbation
$h$ , the required vector is

\begin{equation}
\begin{aligned}\xi^{z}(z,y) & =-\frac{z}{2}\int^{z}_{0}\dd u\,uh_{zz}(u,y),\\
\xi^{a}(z,y) & =-\int^{z}_{0}\dd u\,\gamma^{ab}(u,y)\bigl[u^{2}h_{zb}(u,y)+\partial_{b}\xi^{z}(u,y)\bigr].
\end{aligned}
\label{eq:FGxi}
\end{equation}
The radial FG conditions determine \eqref{eq:FGxi}. Its lower limits
impose $\xi^{z}/z\to0$ and $\xi^{a}\to0$ at the boundary. On the
normalizable source class the integrals converge and preserve $A=0$,
$B=\tau$, while setting $h_{z\mu}=0$. As a diffeomorphism of the
AdS background, the improvement solves the homogeneous linearized
equation and has zero linearized Weyl tensor.

\subsection{Evaluation at an observation point\label{sec:adapted}}

At each observation point $(z,y)$, choose a real flat conformal chart
adapted to the FG normal through $y$, with $\omega(y)=0$, $D_{a}\omega(y)=0$,
and a flat orthonormal tangent basis. Along that normal the Poincare
radial coordinate is $u=z$, the flat tangential coordinate is constant,
and the tangential Jacobi matrix is $E_{a}{}^{c}(z)$, with derivative
$-zP_{a}{}^{c}$.

The FG correction involves the following weighted integral of the
tangential pre-FG perturbation along the boundary normal:

\begin{equation}
J_{cd}(z)=\int^{z}_{0}\dd s\,s^{3}H_{cd}(s,x).\label{eq:J}
\end{equation}

Here $s$ is a dummy Poincare radial coordinate and $x$ is held fixed.
The weight $s^{3}$ follows from the radial FG gauge equations; the
lower limit anchors the diffeomorphism at the boundary. The radial
diffeomorphism itself vanishes on the chosen line, but its tangential
derivatives do not. Evaluating \eqref{eq:FGxi} gives the pointwise
corrected tensor

\begin{equation}
h^{\mathrm{FG}}_{ab}(z,y)=E_{a}{}^{c}E_{b}{}^{d}H_{cd}(z,x)+\frac{2}{z^{2}}E_{(a}{}^{c}P_{b)}{}^{d}J_{cd}(z).\label{eq:FGpoint}
\end{equation}

The two terms in \eqref{eq:FGpoint} are the pulled-back tangential
perturbation and its FG improvement. Appendix \ref{app:FGcompletion}
derives the tangential derivatives of $\xi$ that survive on the observation
line. For $H=u\widehat{\tau}+O(u^{3})$ , one has $J=z^{5}\widehat{\tau}/5+O(z^{7})$
, so the improvement begins at $z^{3}$ and preserves the leading
normalizable coefficient. Since $\partial_{u}I^{(1)}=2uI^{(0)}$ ,

\begin{equation}
J_{cd}(z)=-\frac{c}{2}\bigl[I^{(1)}_{cd}(z)-C_{cd}\bigr].\label{eq:Jendpoint}
\end{equation}

For $C=0$, the Ward identities reduce both terms to one light-cone
cut integral:

\begin{equation}
\begin{aligned}h^{\mathrm{FG}}_{ab}(z,y)=\frac{3}{2\pi z^{2}}\int\dd^{3}x'\,\delta(Q_{z})\Bigl[ & E_{a}{}^{c}E_{b}{}^{d}w_{(c}w^{e}\widehat{\tau}_{d)e}(x')\\
 & -E_{(a}{}^{c}P_{b)}{}^{d}w_{c}w_{d}w^{e}w^{f}\widehat{\tau}_{ef}(x')\Bigr].
\end{aligned}
\label{eq:correctedcut}
\end{equation}

The second line of \eqref{eq:correctedcut} is the curvature-dependent
improvement and vanishes for $P=0$. To act on a stress in its original
representative, apply \eqref{eq:Weyl} with the coordinate and measure
Jacobians. If the transformed flat stress satisfies the stated source
conditions, the paired field obeys the linearized equations, $h_{z\mu}=0$,
and $h_{ab}=z\tau_{ab}+O(z^{3})$.

If a broader source class has a nonzero $C$, the exact gauge-restored
expression includes

\begin{equation}
\Delta h^{\mathrm{FG}}_{ab}=\frac{c}{z^{2}}E_{(a}{}^{c}P_{b)}{}^{d}C_{cd}.\label{eq:extracut}
\end{equation}
This is the lower endpoint of the anchored radial integral. Omitting
it can change the boundary coefficient by $-cP_{(a}{}^{c}C_{b)c}$.
It vanishes for the rapidly decaying timelike sources; a wider source
space requires a compatible endpoint prescription.

\subsection{Example: the global boundary cylinder\label{sec:globalcut}}

In global AdS4 write $f=1+r^{2}$. The Lorentzian metric and its FG
form are

\begin{equation}
\begin{aligned}G & =-f\dd t^{2}+\frac{\dd r^{2}}{f}+r^{2}\dd\Omega^{2},\\
r & =z^{-1}-\frac{z}{4},\qquad G=z^{-2}\left[\dd z^{2}-a^{2}\dd t^{2}+b^{2}\dd\Omega^{2}\right],\\
a & =1+\frac{z^{2}}{4},\qquad b=1-\frac{z^{2}}{4},\qquad0<z<2.
\end{aligned}
\label{eq:globalFG}
\end{equation}

The Einstein cylinder $g_{(0)}=-\dd t^{2}+\dd\Omega^{2}$ is conformally
flat, with $P_{t}{}^{t}=-1/2$, $P_{i}{}^{j}=\delta_{i}{}^{j}/2$,
and $E=\operatorname{diag}(a,b,b)$. Let $\mu=\Omega\cdot\Omega'$,
$\Delta=t'-t$, and

\begin{equation}
S(z;t,\Omega;t',\Omega')=\frac{a}{z}\cos\Delta-\frac{b}{z}\mu.\label{eq:globalS}
\end{equation}
The nearest future and past light-cone cuts satisfy $S=0$, or $\cos\Delta=(b/a)\mu$.
An explicit adapted flat chart on the boundary is

\begin{equation}
\begin{aligned}W & =\cos\Delta+\mu>0,\qquad-\pi<\Delta<\pi,\\
x'^{0} & =\frac{2\sin\Delta}{W},\qquad\boldsymbol{x}'=\frac{2\boldsymbol{\Omega}'_{\perp}}{W},\qquad e^{\omega}=\frac{W}{2}.
\end{aligned}
\label{eq:globalchart}
\end{equation}
Here $\boldsymbol{\Omega}'_{\perp}$ consists of the two components
tangent to the observer's direction $\Omega$. The observation line
has $x=0$. With the tensor coordinate transformation understood,
$\widehat{\tau}=(W/2)(\chi^{-1})^{*}\tau$. The remaining conversion
factors are

\begin{equation}
Q_{z}=\frac{4z}{W}S,\qquad\dd^{3}x'=\left(\frac{2}{W}\right)^{3}\dd t'\dd\Omega',\qquad\delta(Q_{z})=\frac{W}{4z}\delta(S).\label{eq:globalconvert}
\end{equation}

Equations \eqref{eq:correctedcut} and \eqref{eq:globalchart}-{}-\eqref{eq:globalconvert}
give the FG response within this chart, including the Schouten and
source weights. The radial derivative of the light-cone defining function
is

\begin{equation}
\left.\partial_{z}S\right|_{S=0}=\frac{\mu}{1+z^{2}/4},\label{eq:variablecutweight}
\end{equation}
It varies along the light-cone cut and cannot be taken outside the
integral. Its source gradient need not define a conformal-Killing
current. The Ward argument must therefore be applied to the transformed
flat stress, retaining the FG correction.

We take the connected component of $W>0$ containing the observation
boundary point, equivalently $|\Delta|+\arccos\mu<\pi$ on the universal
cover. This formula requires the transformed source to satisfy Section
\ref{sec:spectral}'s falloff and endpoint conditions. For global
periodic modes we instead use the master fields to fix the solution
and its FG improvement, then construct a functional on the complete
light-cone cuts.

\section{Global FG reconstruction on the complete light-cone cuts\label{sec:globaldynamic}}

\subsection{Physical normal modes and their FG improvements\label{sec:globalmodes}}

In this section the boundary stress is a superposition of the regular
scalar and vector mode stresses specified below, with convergence
bounds given in Section \ref{sec:analyticcut}. The corresponding
master fields determine the global linearized Weyl tensor. The radial
integrals below construct the gauge vector that puts the corresponding
metric in normalizable FG form. Their metric contribution is $\Lie_{\xi}G$,
with zero linearized Weyl tensor by \eqref{eq:Weylinvariance}. These
integrals are therefore needed only to express the metric in FG gauge;
the physical curvature can be computed directly from the master fields
or the full pre-FG gauge perturbation.

For the globally regular vacuum problem with fixed boundary metric,
the scalar and vector frequencies are $\omega_{S}=\pm(\ell+1+2n)$
and $\omega_{V}=\pm(\ell+2+2n)$, with $\ell\ge2$, $n\ge0$ \citep{KodamaIshibashi:2003,MichalogiorgakisPufu:2007,Bakas:2009}.
The scalar condition is Neumann in pure AdS, the zero-mass limit of
the Robin condition in \citep{MichalogiorgakisPufu:2007}; the vector
condition is Dirichlet.

Let $\sigma_{ij}$ be the unit-sphere metric and $D_{i}$ its covariant
derivative. We use orthonormal scalar harmonics $Y_{\ell m}$, with
$D^{i}D_{i}Y_{\ell m}=-\lambda_{\ell}Y_{\ell m}$ and $\lambda_{\ell}=\ell(\ell+1)$,
and orthonormal transverse vector harmonics $V^{\ell m}_{i}=\lambda^{-1/2}_{\ell}\epsilon_{i}{}^{j}D_{j}Y_{\ell m}$.
Here $\epsilon_{ij}$ is the sphere area form. Orthonormality uses
the sphere volume measure, with vector indices contracted using $\sigma^{ij}$.
Primes below denote derivatives with respect to $r$.

For a scalar mode of unit density coefficient, the boundary Ward solution
is

\begin{equation}
\begin{aligned}\tau_{tt} & =Y_{\ell m}e^{-i\omega t},\qquad\tau_{ti}=\frac{i\omega}{\lambda_{\ell}}D_{i}Y_{\ell m}e^{-i\omega t},\\
\tau_{ij} & =\left[\frac{\lambda_{\ell}-1-\omega^{2}}{\lambda_{\ell}-2}\sigma_{ij}Y_{\ell m}+\frac{\lambda_{\ell}-2\omega^{2}}{\lambda_{\ell}(\lambda_{\ell}-2)}D_{i}D_{j}Y_{\ell m}\right]e^{-i\omega t}.
\end{aligned}
\label{eq:dynamicWardS}
\end{equation}
For a unit transverse vector coefficient, $\tau_{ti}=V^{\ell m}_{i}e^{-i\omega t}$,
$\tau_{tt}=0$, and $\tau_{ij}=i\omega(D_{i}V_{j}+D_{j}V_{i})e^{-i\omega t}/(\lambda_{\ell}-2)$.
Let $\varrho=\arctan r$, $x=\sin^{2}\varrho$, and define

\begin{equation}
\begin{aligned}F^{S}_{\ell n}(x) & =_{2}F_{1}(-n,n+\ell+1;\ell+3/2;x),\\
F^{V}_{\ell n}(x) & =_{2}F_{1}(-n,n+\ell+2;\ell+3/2;x),\\
\Phi^{S}_{\ell n} & =\frac{6}{\lambda_{\ell}(\lambda_{\ell}-2)}(\sin\varrho)^{\ell+1}\frac{F^{S}_{\ell n}(x)}{F^{S}_{\ell n}(1)},\\
\Phi^{V}_{\ell n} & =-\frac{3}{\lambda_{\ell}-2}(\sin\varrho)^{\ell+1}\cos\varrho\frac{F^{V}_{\ell n}(x)}{F^{V}_{\ell n}(1)}.
\end{aligned}
\label{eq:regularmasters}
\end{equation}

These master profiles solve $(f\Phi')'+(\omega^{2}/f-\lambda_{\ell}/r^{2})\Phi=0$,
normalized by the unit boundary stress. The scalar metric profiles
$H,Q,\mathcal{N}$ follow from the standard harmonic reconstruction
\citep{KodamaIshibashi:2003}; $\alpha,\beta_{S},\vartheta$ determine
its FG improvement:

\begin{equation}
\begin{aligned}H & =\frac{\lambda_{\ell}\Phi^{S}}{2r}+\Phi^{S\prime}-\frac{\omega^{2}r}{f}\Phi^{S},\qquad Q=\frac{\lambda_{\ell}\Phi^{S}}{2r}+f\Phi^{S\prime},\\
\mathcal{N} & =-i\omega\left(\frac{\Phi^{S}}{f}+r\Phi^{S\prime}\right),\\
\alpha & =\frac{\sqrt{f}}{2}\int^{\infty}_{r}\frac{H(u)}{\sqrt{f(u)}}\dd u,\qquad\beta_{S}=\int^{\infty}_{r}\frac{\alpha(u)}{u^{2}f(u)}\dd u,\\
\vartheta & =-\int^{\infty}_{r}\left[\frac{\mathcal{N}(u)}{f(u)}-\frac{i\omega\alpha(u)}{f(u)^{2}}\right]\dd u.
\end{aligned}
\label{eq:dynamicFGS}
\end{equation}

The improvement vector has components $\xi^{r}=\alpha Y_{\ell m}$,
$\xi^{i}=\beta_{S}D^{i}Y_{\ell m}$, and $\xi^{t}=\vartheta Y_{\ell m}$,
with the common time dependence. Its Lie derivative removes the radial
components of the pre-FG gauge perturbation. With $N=(\ell,m,n,\operatorname{sign}\omega)$,
the unit FG response is

\begin{equation}
\begin{aligned}\mathcal{H}^{S,N}_{tt} & =(fH-2r\alpha+2i\omega f\vartheta)Y_{\ell m},\\
\mathcal{H}^{S,N}_{ti} & =(-f\vartheta-i\omega r^{2}\beta_{S})D_{i}Y_{\ell m},\\
\mathcal{H}^{S,N}_{ij} & =(r^{2}Q+2r\alpha)\sigma_{ij}Y_{\ell m}+2r^{2}\beta_{S}D_{i}D_{j}Y_{\ell m}.
\end{aligned}
\label{eq:globalHscalar}
\end{equation}

The vector sector follows the same pattern. Its physical representative
has a tangential profile $f(r\Phi^{V})'$ and a radial component removed
by $\xi^{i}=\eta_{V}V^{i}$, where $\eta_{V}=-i\omega\int^{\infty}_{r}\Phi^{V}(u)/(uf(u))\dd u$.
The unit FG response is

\begin{equation}
\begin{aligned}\mathcal{H}^{V,N}_{ti} & =\left[f(r\Phi^{V})'-i\omega r^{2}\eta_{V}\right]V^{\ell m}_{i},\\
\mathcal{H}^{V,N}_{ij} & =r^{2}\eta_{V}(D_{i}V^{\ell m}_{j}+D_{j}V^{\ell m}_{i}),\qquad\mathcal{H}^{V,N}_{tt}=0.
\end{aligned}
\label{eq:globalHvector}
\end{equation}
The normalizations can be checked directly at large radius:

\begin{equation}
\begin{aligned}\Phi^{S} & =\frac{6}{\lambda_{\ell}(\lambda_{\ell}-2)}\left[1+\frac{\lambda_{\ell}-\omega^{2}}{2r^{2}}+O(r^{-4})\right],\\
\Phi^{V} & =-\frac{3}{(\lambda_{\ell}-2)r}\left[1+\frac{\lambda_{\ell}-\omega^{2}-2}{6r^{2}}+O(r^{-4})\right].
\end{aligned}
\label{eq:dynamicboundarycheck}
\end{equation}
The improvement cancels the coordinate terms proportional to $r$.
The scalar limits are $r\mathcal{H}^{S,N}_{tt}\to Y_{\ell m}$, $r\mathcal{H}^{S,N}_{ti}\to i\omega D_{i}Y_{\ell m}/\lambda_{\ell}$,
and \eqref{eq:dynamicWardS} for the spatial components. For vectors,
$r\mathcal{H}^{V,N}_{ti}\to V^{\ell m}_{i}$ and $r\mathcal{H}^{V,N}_{ij}\to i\omega(D_{i}V_{j}+D_{j}V_{i})/(\lambda_{\ell}-2)$.
Since $rz\to1$, both give $h_{ab}/z\to\tau_{ab}$ and $A=0$.

These responses have zero radial components. Using $\rho=\tau_{tt}$
and $j_{i}=\tau_{ti}$, define the spectral kernels below. The vector
harmonic pairing selects the transverse part of $j$.

\begin{equation}
\begin{aligned}\mathscr{G}^{\rho}_{ab}(r;\Omega;s,\Omega') & =\frac{1}{2\pi}\sum_{N\in S}\mathcal{H}^{S,N}_{ab}(r,\Omega)e^{i\omega_{N}s}Y^{*}_{\ell m}(\Omega'),\\
\mathscr{G}^{j'}_{ab}(r;\Omega;s,\Omega') & =\frac{1}{2\pi}\sum_{N\in V}\mathcal{H}^{V,N}_{ab}(r,\Omega)e^{i\omega_{N}s}V^{*j'}_{\ell m}(\Omega').
\end{aligned}
\label{eq:globalG}
\end{equation}

At fixed $\ell$, the ratios $F^{S,V}(x)/F^{S,V}(1)$ are Jacobi-polynomial
ratios $P^{(\ell+1/2,\mp1/2)}_{n}(1-2x)/P^{(\ell+1/2,\mp1/2)}_{n}(-1)$
\citep{DLMF}. Their radial derivatives grow polynomially in $n$
on compact positive-radius intervals, as do the anchored gauge integrals.
A finite angular cutoff therefore defines ordinary periodic temporal
distributions in \eqref{eq:globalG}. Pairing over a period extracts
the normal-mode coefficients and gives a solution with $A=0$, $B=\tau$,
built from regular pre-FG gauge perturbations. FG coordinates can
themselves be singular at the center, where the boundary normals form
a caustic.

\subsection{Radial information in the global FG improvement\label{sec:nogo}}

Define the complete nearest light-cone cuts of $X=(t,r,\Omega)$ by

\begin{equation}
t'_{\sigma}(\Omega')=t+\sigma\alpha_{X}(\Omega'),\qquad\alpha_{X}(\Omega')=\arccos\left(\frac{r}{\sqrt{1+r^{2}}}\Omega\cdot\Omega'\right),\qquad\sigma=\pm1.\label{eq:completecuts}
\end{equation}
Put $\beta(r)=\arctan(1/r)$ , so $\beta\le\alpha_{X}\le\pi-\beta$
. The first example in Appendix \ref{app:globalexamples} constructs
a smooth conserved, traceless stress from the regular scalar frequencies
$\omega=\pm(3+2n)$ , with $A=0$ . When $\beta(r)$ exceeds the pulse
half-width $\delta$, the stress vanishes near both complete light-cone
cuts, yet the boundary-anchored global FG metric has a nonzero angular
component.

In this example the master field, pre-FG gauge perturbation, and linearized
Weyl tensor vanish near the observation point. The surviving FG component
comes from the improvement integrated along outer portions of the
boundary-normal curves. Any ordinary distribution supported on the
complete light-cone cuts annihilates the stress, since it vanishes
in a neighborhood of both. Such a kernel therefore cannot reproduce
the FG metric for every smooth normal-mode source.

The example also obeys $\mathfrak{A}^{*}\tau=-\tau$, where $\mathfrak{A}(t,\Omega)=(t+\pi,-\Omega)$.
Normalizability and this antipodal relation do not remove the radial
improvement. The spectral kernel \eqref{eq:globalG} still reconstructs
the field; an ordinary distribution supported only on the complete
light-cone cuts cannot reproduce the FG metric for every smooth source.

The second example in Appendix \ref{app:globalexamples} applies unmodified
antipodal gluing to the lowest regular scalar mode, $\ell=2,\omega=3$
. It produces a nonzero boundary-metric coefficient from a mode whose
global solution has $A=0$ . Even a single-frequency restriction therefore
fails to justify that weight. On a finite mode space one may construct
other ordinary light-cone cut kernels when the restricted modes are
linearly independent there, but their weights depend on the chosen
space.

\subsection{A convergent series of contact terms on the complete light-cone cuts\label{sec:analyticcut}}

To represent the full FG response on the light-cone cuts, Taylor-expand
the source in the spectral pairing. At finite angular bandwidth, a
sufficient analyticity condition is $\sum_{N}e^{a|\omega_{N}|}|b_{N}|<\infty$,
with $a>\pi$, for scalar or transverse-vector coefficients $b_{N}$.
Choose $0<\epsilon<a-\pi$ and a smooth weight $\chi$, supported
in $(-\pi-\epsilon,\pi+\epsilon)$, such that

\begin{equation}
\sum_{k\in\mathbb{Z}}\chi(s+2\pi k)=1.\label{eq:periodweight}
\end{equation}
For each source direction put $s_{\sigma}=\sigma\alpha_{X}(\Omega')$
and center the period window on that light-cone cut. Periodicity and
\eqref{eq:periodweight} give window-independent summed pairings.
Define the moments

\begin{equation}
C^{I}_{ab,q,\sigma}(r;\Omega,\Omega')=\frac{1}{q!}\left\langle \mathscr{G}^{I}_{ab}(r;\Omega;s+s_{\sigma},\Omega'),\chi(s)s^{q}\right\rangle _{s},\qquad I=\rho,j'.\label{eq:cutmoments}
\end{equation}

Taylor expansion uses only real time derivatives of the supplied real-coordinate
source. Averaging the two equal expansions gives the formula over
the complete light-cone cuts

\begin{equation}
\begin{aligned}h^{{\rm FG}}_{ab}(X)=\frac{1}{2}\sum_{\sigma=\pm1}\sum^{\infty}_{q=0}\int_{S^{2}}\dd\Omega'\Bigl[ & C^{\rho}_{ab,q,\sigma}\partial^{q}_{t}\rho(t+s_{\sigma},\Omega')\\
 & +C^{j'}_{ab,q,\sigma}\partial^{q}_{t}j_{j'}(t+s_{\sigma},\Omega')\Bigr].
\end{aligned}
\label{eq:analyticfullcut}
\end{equation}
Equivalently, the kernel is $\frac{1}{2}\sum_{\sigma,q}(-1)^{q}C^{I}_{ab,q,\sigma}\partial^{q}_{t'}\delta(t'-t-s_{\sigma})$.
Each contact term lies on a complete future or past light-cone cut,
with coefficients fixed by the global mode sum and FG improvement.
The full series defines a generalized kernel on the time-analytic
normal-mode sources satisfying the stated bounds. It need not converge
on an arbitrary smooth source.

On the pairing support, $|s|<\pi+\epsilon<a$. Choose a strip width
$a_{0}$ with $\pi+\epsilon<a_{0}<a$. Cauchy estimates bound the
input derivatives by $Cq!a^{-q}_{0}$. The temporal distribution contributes
only polynomial factors in $q$, so the moments converge geometrically
and justify the Taylor interchange. For infinitely many angular modes,
a stronger sufficient condition is

\begin{equation}
\sum_{N}e^{b\omega^{2}_{N}}|b_{N}|<\infty,\qquad b>0.\label{eq:Gaussianmodes}
\end{equation}

With $N_{*}=1+\ell+n$, the finite hypergeometric sums, their nonzero
boundary denominators, differentiated harmonics, and radial gauge
integrals give the crude bound $\|\mathcal{H}^{N}\|_{C^{k}(J)}\le\exp[C_{J,k}N_{*}\log(N_{*}+2)]$
on each compact bulk region $J$ with $r>0$. Since $|\omega_{N}|$
is comparable to $N_{*}$, the Gaussian weight dominates this bound
and all fixed time and spatial derivatives. Define the full kernel
by first summing the Taylor series at each angular cutoff and then
taking that cutoff to infinity. The resulting paired field agrees
mode by mode with \eqref{eq:globalG}; its derivatives converge on
compact bulk regions. Finite normal-mode sums are an elementary admissible
example. We do not exchange these two limits without an additional
uniform bound.

For the time-analytic normal-mode sources satisfying these bounds,
the derivatives determine source values beyond a neighborhood of the
light-cone cuts and retain the radial information in Section \ref{sec:nogo}.
The counterexample built from a smooth pulse vanishes on an open time
interval, which a nonzero analytic source cannot do. Changing $\chi$
changes the moments but leaves their summed action unchanged.

\section{General spacelike momentum in the Poincare patch\label{sec:spacelikeflat}}

\subsection{Source domain and the normalizable branch\label{sec:spacelikeregime}}

For the flat-frame formulas below, we suppress hats on the stress.
The normalizability restriction in Section \ref{sec:spectral} excludes
the open spacelike cone $p^{2}=|\boldsymbol{k}|^{2}-\omega^{2}>0$
from the original Lorentzian kernel. This section reconstructs that
sector with $A=0$, $B=\tau$, allowing arbitrary frequency subject
to spacelike support and the convergence bounds below. The boundary
Ward identities remain $\partial^{a}\tau_{ab}=0$ and $\eta^{ab}\tau_{ab}=0$.

The radial profiles are standard consequences of the spin-two equation
\citep{kabat,BhattacharjeeMukherjee:2025}. The issue addressed here
is their realization as kernels on real boundary data, with an explicit
source space and an FG completion in curved representatives. Exponential
amplification of spacelike momentum is the familiar obstruction associated
with evanescent modes \citep{ReyRosenhaus:2014}; it requires a stronger
source domain than smoothness alone.

Put $\psi_{ab}=z^{2}H_{ab}$. For a Fourier mode with $q=\sqrt{|\boldsymbol{k}|^{2}-\omega^{2}}>0$,
the flat FG equation and boundary conditions are

\begin{equation}
\psi_{ab}''-\frac{2}{z}\psi_{ab}'-q^{2}\psi_{ab}=0,\qquad\psi_{ab}=z^{3}\widetilde{\tau}_{ab}+O(z^{5}),\qquad A_{ab}=0.\label{eq:spacelikeODE}
\end{equation}

They select the pure normalizable solution

\begin{equation}
\widetilde{H}_{ab}(z,p)=F_{B}(z,q)\widetilde{\tau}_{ab}(p),\qquad F_{B}(z,q)=\frac{3}{z^{2}q^{3}}\bigl[qz\cosh(qz)-\sinh(qz)\bigr].\label{eq:FB}
\end{equation}

The apparent singularity at $q=0$ is removable. Equivalently, using
\citep{DLMF},

\begin{equation}
\begin{aligned}F_{B}(z,q) & =z\,{}_{0}F_{1}\left(;\frac{5}{2};\frac{z^{2}q^{2}}{4}\right)=z\,\Gamma\left(\frac{5}{2}\right)\left(\frac{2}{qz}\right)^{3/2}I_{3/2}(qz),\\
F_{B}(z,q) & =z+\frac{q^{2}z^{3}}{10}+\frac{q^{4}z^{5}}{280}+O(z^{7}).
\end{aligned}
\label{eq:FBbessel}
\end{equation}

Here $I_{3/2}$ is the modified Bessel function of the first kind;
the symbol $K_{z}$ below names a kernel, not the Bessel function
$K_{\nu}$. The entire dependence on $q^{2}$ also gives a continuation
to timelike momentum, with $I_{3/2}$ replaced by the normalized $J_{3/2}$,
agreeing with \eqref{eq:timelike}.

For spacelike momentum, the relation to the original Lorentzian representative
\eqref{eq:spacelike} is

\begin{equation}
F_{B}(z,q)=\frac{3(1+qz)e^{-qz}}{q^{3}z^{2}}-\frac{3}{q^{3}z^{2}}\bigl[\cosh(qz)-qz\sinh(qz)\bigr].\label{eq:subtractA}
\end{equation}

The second term removes the complete boundary-metric branch and its
descendants. This changes the physical boundary data and cannot be
supplied by a boundary-preserving gauge improvement. The selected
solution is already in flat FG gauge. Its growth $F_{B}\sim3e^{qz}/(2zq^{2})$
explains the convergence requirement below; no condition of regularity
at the Poincare horizon is imposed.

\subsection{A real position-space generalized function\label{sec:spacelikeposition}}

Let $\mathcal{S}_{+}=\{(\omega,\boldsymbol{k}):|\boldsymbol{k}|^{2}-\omega^{2}\ge0\}$
and $P=\sqrt{\omega^{2}+|\boldsymbol{k}|^{2}}$. For $a>0$, define
a source space $\cA^{{\rm sp}}_{a}$ by support in $\mathcal{S}_{+}$
and the weighted bounds

\begin{equation}
\|\tau\|_{a,N}:=\sum_{b,c}\int_{\mathcal{S}_{+}}e^{a\sqrt{|\boldsymbol{k}|^{2}-\omega^{2}}}(1+P)^{N}|\widetilde{\tau}_{bc}(\omega,\boldsymbol{k})|\,\dd\omega\,\dd^{2}k<\infty,\qquad N=0,1,2,\ldots.\label{eq:spacelikenorm}
\end{equation}

Finite Fourier measures are also allowed: the integral in \eqref{eq:spacelikenorm}
is then taken over the absolute weight of the modes. This includes
delta functions in momentum space, such as those describing plane
waves or sources independent of time, whose modes are supported at
$\omega=0$. The polynomial weights control differentiation, including
large frequencies near the null cone where $q$ alone does not control
$P$.

To define a position-space representative, extend $F_{B}$ by its
entire series in $p^{2}$. Choose a smooth cutoff $\zeta(P/Q)$, equal
to one for $P\le Q$ and zero for $P\ge2Q$, and put

\begin{equation}
\begin{aligned}K^{(Q)}_{z}(w) & =\int\frac{\dd\omega\,\dd^{2}k}{(2\pi)^{3}}\,\zeta(P/Q)e^{-i\omega w^{0}+i\boldsymbol{k}\cdot\boldsymbol{w}}F_{B}\left(z,\sqrt{|\boldsymbol{k}|^{2}-\omega^{2}}\right),\\
H_{ab}(z,x) & =(K_{z}*\tau_{ab})(x):=\lim_{Q\to\infty}(K^{(Q)}_{z}*\tau_{ab})(x).
\end{aligned}
\label{eq:spacelikepairing}
\end{equation}

At finite cutoff the kernel is a smooth, rapidly decreasing function
of real $w$, so its convolution with the bounded source is an ordinary
real integral. For $0<z<a$, the limit is a continuous generalized
function on $\cA^{{\rm sp}}_{a}$. All fixed bulk derivatives of the
paired field converge on compact bulk regions. Its action on this
source space is independent of the chosen continuation away from $\mathcal{S}_{+}$.

The same position-space kernel can be written as a series of contact
terms at coincident boundary points:

\begin{equation}
\begin{aligned}K_{z}(w) & =6\sum^{\infty}_{n=0}\frac{(n+1)z^{2n+1}}{(2n+3)!}(-\Box_{\eta})^{n}\delta^{(3)}(w)\\
 & =z\,{}_{0}F_{1}\left(;\frac{5}{2};-\frac{z^{2}}{4}\Box_{\eta}\right)\delta^{(3)}(w),\qquad\Box_{\eta}=-\partial^{2}_{t}+\partial^{2}_{x}+\partial^{2}_{y}.
\end{aligned}
\label{eq:spacelikeseries}
\end{equation}

Integrating these contact terms against the stress evaluates its value
and derivatives at the observation boundary point. Summing the terms
gives

\begin{equation}
H_{ab}(z,x)=z\tau_{ab}(x)-\frac{z^{3}}{10}\Box_{\eta}\tau_{ab}(x)+\frac{z^{5}}{280}\Box^{2}_{\eta}\tau_{ab}(x)-\cdots.\label{eq:spacelikeresponse}
\end{equation}

Equation \eqref{eq:spacelikeseries} means that the series converges
when applied to sources obeying the stated bounds. Each contact term
takes a derivative of finite order, while the full kernel may require
derivatives of all orders. The sum does not, in general, define an
ordinary distribution acting on every smooth compactly supported source,
so its support cannot be inferred by treating it as a finite list
of contact terms. Only real-coordinate values and derivatives of the
stress are used.

A sufficient condition stated entirely in real position space is

\begin{equation}
\bigl\|\partial^{\alpha}(-\Box_{\eta})^{n}\tau_{ab}\bigr\|_{\infty}\le C_{\alpha,ab}\frac{(2n)!}{a^{2n}},\qquad n=0,1,2,\ldots,\qquad a>z.\label{eq:spacelikederivativebound}
\end{equation}

Taking $\alpha=0$ suffices for the field; every fixed $\alpha$ permits
its corresponding derivatives. Equation \eqref{eq:spacelikenorm}
implies these estimates. Conversely, \eqref{eq:spacelikederivativebound}
directly defines an additional domain for the entire derivative-series
kernel, without a Fourier support assumption. It controls powers of
the wave operator and need not imply analyticity in each boundary
coordinate. This extension is useful under conformal changes of frame.
Appendix \ref{app:spacelikeconvergence} proves convergence and the
radial equation. A finite $a$ gives reconstruction for $0<z<a$;
imposing the bounds for every finite $a>0$ suffices at every finite
Poincare depth.

\subsection{Nonzero frequency and the zero-frequency limit\label{sec:zerofrequency}}

A simple two-polarization example makes the allowed time dependence
explicit. Let $k>0$, $v=\omega/k$, $|v|<1$, and take

\begin{equation}
\tau_{ab}(t,x,y)=\operatorname{Re}\left[e^{-i\omega t+ikx}\begin{pmatrix}M & -vM & N\\
-vM & v^{2}M & -vN\\
N & -vN & (1-v^{2})M
\end{pmatrix}_{ab}\right].\label{eq:spacelikeexample}
\end{equation}

Both constants $M,N$ may be complex. The matrix is Lorentzian traceless
and obeys $\omega\widetilde{\tau}_{0b}+k\widetilde{\tau}_{1b}=0$,
so conservation holds. Multiplication by $F_{B}(z,k\sqrt{1-v^{2}})$
gives its normalizable FG response. Finite mode sums satisfy \eqref{eq:spacelikenorm}
in its measure form for every $a$.

At fixed spatial momentum, $\omega\to0$ gives $q\to|\boldsymbol{k}|$,
so the same kernel yields the zero-frequency response with $-\Box_{\eta}\to-\Delta_{\mathbb{R}^{2}}$.
This is a limit within the spacelike sector, not a separate reconstruction
problem. More generally, suppose the Fourier measures converge when
integrated against every smooth function of momentum with compact
support. Uniform versions of \eqref{eq:spacelikenorm} at some $a>z$
then give convergence of the paired fields and their fixed derivatives
on compact bulk regions inside $0<z<a$. A measure supported at $\omega=0$
is allowed. The mode at $q=0$ has multiplier $z$.

A Hadamard finite part may extend a divergent regulated integral,
but the boundary conditions $A=0$, $B=\tau$ must still be specified.
Here \eqref{eq:spacelikeODE} and \eqref{eq:subtractA} fix the branch
before pairing; an odd-power selection is not an independent prescription.

\subsection{Comparison with complex-coordinate smearing\label{sec:spacelikecomplex}}

There is a modewise comparison with Eq. (37) of \citep{kabat}. For
a mode $e^{-i\omega t+i\boldsymbol{k}\cdot\boldsymbol{x}}$, its complex-ball
integral gives

\begin{equation}
\begin{aligned}\frac{3}{4\pi z^{2}}\int_{s^{2}+|\boldsymbol{v}|^{2}<z^{2}}\dd s\,\dd^{2}v\,e^{-i\omega s-\boldsymbol{k}\cdot\boldsymbol{v}} & =F_{B}(z,q),\\
q^{2} & =|\boldsymbol{k}|^{2}-\omega^{2}.
\end{aligned}
\label{eq:complexballcomparison}
\end{equation}

Rotational ball moments give an entire function of $|\boldsymbol{k}|^{2}-\omega^{2}$,
namely \eqref{eq:FBbessel}. A literal complex-coordinate smearing
of a general source requires the relevant holomorphic spatial continuation;
a stronger sufficient Fourier bound is $\int e^{aP}|\widetilde{\tau}|\dd^{3}p<\infty$,
with $a>z$. Our spacelike bound controls $q$, not $P$, and need
not supply that continuation near the null cone. Thus the mode profiles
agree, while \eqref{eq:spacelikepairing} and \eqref{eq:spacelikeseries}
specify a reconstruction directly from real data on their stated domains. 

\section{FG completion and a global position-space kernel\label{sec:spacelikeglobal}}

\subsection{The weighted radial integral and its kernel\label{sec:spacelikeimprovement}}

The gauge-independent curvature can be calculated directly from $H=K_{z}*\widehat{\tau}$.
To format the result in the FG coordinates of a curved conformally
flat representative, use the boundary-anchored improvement of Section
\ref{sec:FGprojection}. The corresponding weighted radial integral
is generated by the following series of contact terms:

\begin{equation}
\begin{aligned}L_{z}(w) & :=\int^{z}_{0}\dd s\,s^{3}K_{s}(w)\\
 & =6\sum^{\infty}_{n=0}\frac{(n+1)z^{2n+5}}{(2n+5)(2n+3)!}(-\Box_{\eta})^{n}\delta^{(3)}(w)\\
 & =\frac{z^{5}}{5}\,{}_{0}F_{1}\left(;\frac{7}{2};-\frac{z^{2}}{4}\Box_{\eta}\right)\delta^{(3)}(w).
\end{aligned}
\label{eq:spacelikeL}
\end{equation}
On a spacelike mode its multiplier is

\begin{equation}
\widetilde{L}_{z}(p)=\frac{3}{q^{5}}\left[\bigl((qz)^{2}+3\bigr)\sinh(qz)-3qz\cosh(qz)\right],\qquad\widetilde{L}_{z}(0)=\frac{z^{5}}{5}.\label{eq:spacelikeLmultiplier}
\end{equation}

At an observation point $(z,y)$, choose the adapted real conformal
frame of Section \ref{sec:adapted}, with $x_{0}=\chi(y)$, $\omega(y)=0$,
and $D\omega(y)=0$. Transform the stress by \eqref{eq:Weyl}. The
full FG kernel acting on its flat components is

\begin{equation}
\begin{aligned}\mathscr{K}_{ab}{}^{cd}(z,y;x')={} & E_{a}{}^{(c}E_{b}{}^{d)}K_{z}(x_{0}-x')\\
 & +\frac{2}{z^{2}}E_{(a}{}^{(c}P_{b)}{}^{d)}L_{z}(x_{0}-x'),\\
h^{{\rm FG}}_{ab}(z,y) & =\int\dd^{3}x'\,\mathscr{K}_{ab}{}^{cd}(z,y;x')\widehat{\tau}_{cd}(x'),\\
h^{{\rm FG}}_{z\mu} & =0.
\end{aligned}
\label{eq:spaceliketensorkernel}
\end{equation}
The second line of the kernel is the improvement. Since $L_{z}*\widehat{\tau}=z^{5}\widehat{\tau}/5+O(z^{7})$,
it begins at order $z^{3}$, preserves $A=0$, $B=\tau$, and leaves
the linearized Weyl tensor unchanged. Both kernels converge under
the same sufficient bounds. Their source-variable differential operators
act before multiplication by the observation-point tensors $E,P$.

\subsection{Covariant contact terms in curved representatives\label{sec:globalcontacts}}

A boundary conformal transformation need not preserve spacelike Fourier
support or its convergence bounds. Consequently the flat-frame condition
$p^{2}>0$ should not be imposed as a coordinate-independent spectral
restriction on the cylinder. For a general locally conformally flat
representative, the position-space formulation uses covariant derivatives
of the boundary stress, obtained by transforming to adapted flat frames.
The source bounds must be checked in that representative.

At each $y$, let $\mathcal{T}_{y}$ denote the actual Weyl and tensor
transformation to an adapted flat frame. Hold this frame fixed while
differentiating its source variable, then evaluate at $x_{0}$ and
convert the output indices back to the tangent basis at $y$. Define

\begin{equation}
(\mathcal{D}_{n}\tau)_{ab}(y):=\left[\mathcal{T}^{-1}_{y}(-\Box_{\eta})^{n}\mathcal{T}_{y}\tau\right]_{ab}(y),\qquad\mathcal{D}_{0}=1.\label{eq:covariantderivatives}
\end{equation}

These are differential operators of order $2n$, including the tensor
and Weyl factors. For example,

\begin{equation}
(\mathcal{D}_{1}\tau)_{ab}=-D^{c}D_{c}\tau_{ab}+3P_{c}{}^{c}\tau_{ab}.\label{eq:covariantD1}
\end{equation}

The kernel therefore includes curvature terms beyond replacing $\Box_{\eta}$
by the curved wave operator. With all indices now in the original
representative, \eqref{eq:spaceliketensorkernel} becomes

\begin{equation}
\begin{aligned}h^{{\rm FG}}_{ab}(z,y)={} & 6\sum^{\infty}_{n=0}\frac{(n+1)z^{2n+1}}{(2n+3)!}\\
 & \quad\times\left[E_{a}{}^{c}E_{b}{}^{d}+\frac{2z^{2}}{2n+5}E_{(a}{}^{c}P_{b)}{}^{d}\right](\mathcal{D}_{n}\tau)_{cd}(y).
\end{aligned}
\label{eq:globalFGseries}
\end{equation}

Let $\boldsymbol{\delta}(y,y')$ be the contact term at $y=y'$, with
tensor indices arranged so that integrating it against the boundary
stress returns that stress at $y$. The integration uses the boundary
volume measure. The full position-space kernel is then a series of
covariant contact terms:

\begin{equation}
\begin{aligned}\mathscr{K}_{ab}{}^{c'd'}(z;y,y')={} & 6\sum^{\infty}_{n=0}\frac{(n+1)z^{2n+1}}{(2n+3)!}\\
 & \quad\times\left[E_{a}{}^{e}E_{b}{}^{f}+\frac{2z^{2}}{2n+5}E_{(a}{}^{e}P_{b)}{}^{f}\right]\\
 & \quad\times(\mathcal{D}^{y}_{n}\boldsymbol{\delta})_{ef}{}^{c'd'}(y,y').
\end{aligned}
\label{eq:globalFGcontactkernel}
\end{equation}

Its paired action is \eqref{eq:globalFGseries}. A sufficient convergence
condition, imposed uniformly on each compact boundary set $U$ , is

\begin{equation}
\|D^{\alpha}\mathcal{D}_{n}\tau\|_{U}\le C_{U,\alpha}\frac{(2n)!}{a^{2n}},\qquad n\ge0,\qquad z<a,\label{eq:globalderivativebound}
\end{equation}

Here the bound uses any fixed positive component norm. For a general
locally conformally flat representative, \eqref{eq:globalFGcontactkernel}
gives the FG response on a regular coordinate neighborhood where \eqref{eq:globalderivativebound}
holds. Convergent local expressions agree on overlaps, as shown in
Appendix \ref{app:spacelikeconvergence}. A global metric additionally
requires compatible boundary data and such neighborhoods covering
the bulk region under consideration.

On the Einstein cylinder, $P=\operatorname{diag}(-1,1,1)/2$ and $E=\operatorname{diag}(1+z^{2}/4,1-z^{2}/4,1-z^{2}/4)$
, in an orthonormal tangent basis. Equations \eqref{eq:globalFGcontactkernel}
and \eqref{eq:globalderivativebound} therefore give an explicit generalized
kernel for $0<z<2$ wherever the source bound reaches that depth.
This is a convergence condition on cylinder data, including their
time derivatives. It is not inferred from admissibility in one flat
chart.

For clarity, the first curvature correction is

\begin{equation}
\begin{aligned}h^{{\rm FG}}_{ab}={} & z\tau_{ab}+\frac{z^{3}}{10}\Bigl[-D^{c}D_{c}\tau_{ab}+3P_{c}{}^{c}\tau_{ab}\\
 & \qquad-3P_{a}{}^{c}\tau_{cb}-3P_{b}{}^{c}\tau_{ac}\Bigr]+O(z^{5}).
\end{aligned}
\label{eq:globalFGleading}
\end{equation}

Agreement on overlaps follows from the uniquely determined, convergent
FG expansion with $A=0$ and $B=\tau$. This construction yields the
improved metric; it does not imply that the original patch light-cone
weights can be glued without modification. The two examples in Appendix
\ref{app:globalexamples} illustrate the distinct limitations.

Normalizability and these source bounds do not impose center regularity
or admit arbitrary smooth global data. Time-independent cylinder sources
are included when they satisfy \eqref{eq:globalderivativebound};
their transformed flat stresses generally depend on flat time.

\section{Flat-space limits and celestial holography\label{sec:celestial}}

\subsection{Boundary stress and the soft gravitational sector\label{sec:softstress}}

In asymptotically flat gravity, the leading soft theorem is a BMS
supertranslation Ward identity \citep{HeLysovMitraStrominger:2015};
the subleading theorem gives a celestial stress tensor with Virasoro
Ward identities \citep{KapecMitraRaclariuStrominger:2017}. Conformally
soft wavefunctions describe the associated symmetry generators \citep{DonnayPuhmStrominger:2019}.
See \citep{Pasterski:2021aa} for a review.

The real Lorentzian construction of \citep{Lowe:2024reconstruction}
was already motivated by these connections. Its example uses shockwaves
with nontrivial extended-BMS charges in the large-radius limit \citep{Lowe:2020qan}.
Within AdS/CFT, soft photon theorems follow from CFT Ward identities
\citep{Hijano:2020szl}, while leading and subleading conformally
soft graviton relations follow from shadow stress-tensor identities
in shrinking cylinder time strips \citep{deGioiaRaclariu:2024}. At
null infinity, a Carrollian stress tensor encodes both soft graviton
theorems \citep{BagchiDhivakarDutta:2026}, and flat limits of AdS
correlators give Carrollian amplitudes \citep{AldayNocchiRuzziconiSrikant:2025}.

The spacelike kernels enlarge the classical source family available
for studying these limits. They reconstruct modes with $0<|\omega|<|\boldsymbol{k}|$
and include $\omega=0$ in the same normalized boundary-to-bulk map.
The uniform source bounds of Section \ref{sec:zerofrequency} allow
the zero-frequency limit to be taken within that map. This makes it
possible to compare a slowly varying field with its zero-frequency
endpoint while retaining its curvature and FG improvement.

\subsection{Radius scaling and the zero-frequency limit\label{sec:softscaling}}

Restore the AdS radius $L$, with global physical coordinates $T=Lt$,
$R=Lr$. The metric and regular-mode frequencies become

\begin{equation}
\begin{aligned}G_{L} & =-\left(1+\frac{R^{2}}{L^{2}}\right)\dd T^{2}+\frac{\dd R^{2}}{1+R^{2}/L^{2}}+R^{2}\dd\Omega^{2},\\
\omega^{\mathrm{phys}}_{S} & =\frac{\ell+1+2n}{L},\qquad\omega^{\mathrm{phys}}_{V}=\frac{\ell+2+2n}{L}.
\end{aligned}
\label{eq:softfrequencies}
\end{equation}

Fixed $\ell,n$ give physical frequencies tending to zero as $L\to\infty$,
whereas finite nonzero frequencies require mode numbers growing with
$L$. Finite sums of both polarizations and the analytic packets of
Section \ref{sec:analyticcut} provide candidate low-frequency families.
The general spacelike construction supplies additional exterior families,
including continuously variable boundary frequency.

For a flat boundary frame rescaled in the same way, the kinematic
labels obey

\begin{equation}
\Omega=\frac{\omega}{L},\qquad\boldsymbol{K}=\frac{\boldsymbol{k}}{L},\qquad Q=\frac{q}{L},\qquad Q^{2}=|\boldsymbol{K}|^{2}-\Omega^{2}.\label{eq:softspacelikescaling}
\end{equation}

At fixed nonzero spatial momentum, $\omega\to0$ is the zero-frequency
limit of an exterior field. This alone does not identify an on-shell
radiative soft graviton. Its limiting data may include Coulombic fields,
asymptotic symmetry data, or a radiative soft component, depending
on the stress family and radius scaling. A soft-graviton identification
requires the simultaneous large-radius and low-physical-frequency
limit, together with the corresponding spatial and radial scaling.
In particular, simply setting $\omega=0$ does not establish a BMS
Ward identity or a celestial conformally soft theorem.

A concrete test is to choose $\tau^{(L)}$ and its amplitude scaling,
retain uniform control of the kernel pairings in the bulk region being
scaled, and calculate the limiting Bondi shear, news, memory, and
charges. The holographic normalization must be restored with the chosen
physical boundary metric. For fixed physical $R$, the global limit
probes $r=R/L\to0$, so convergence on a fixed radial region of AdS
is not by itself a uniform flat-limit estimate. Gauge-invariant curvature
or master fields give a direct route to the physical limit; the metric
can then be compared in the asymptotic gauge used to define charges.

\subsection{Improvements and asymptotic symmetry data\label{sec:softgauge}}

At fixed AdS radius, the anchored improvements preserve the boundary
stress and have zero local linearized Weyl tensor. Their flat limits
may nevertheless act nontrivially at null infinity, where a diffeomorphism
can carry a charge or change the asymptotic vacuum \citep{HeLysovMitraStrominger:2015,DonnayPuhmStrominger:2019}.
A limiting calculation must therefore track the improvement's asymptotic
behavior as well as the curvature.

\section{Discussion\label{sec:discussion}}

The real light-cone kernel of \citep{Lowe:2024reconstruction} remains
the simplest route to the physical curvature on its admissible source
class. We have specified the spectral and endpoint conditions that
keep the boundary metric fixed and derived the FG improvement in curved
conformal representatives. Local and global improvements preserve
the prescribed stress and contribute no linearized Weyl curvature,
allowing the curvature to be calculated before imposing FG gauge.

Appendix \ref{app:globalexamples} establishes two distinct limitations.
An ordinary distribution confined to the complete light-cone cuts
cannot recover the FG metric for every smooth global source; the missing
term in the example is pure gauge and has zero local Weyl curvature.
Even an analytic single mode does not justify unmodified antipodal
gluing of the patch weight. Time-analytic superpositions of the regular
global modes satisfying the bounds of Section \ref{sec:analyticcut}
admit the complete-cut representation \eqref{eq:analyticfullcut},
including the full anchored improvement.

The complementary open spacelike sector is reconstructed by \eqref{eq:spacelikepairing}
and \eqref{eq:spacelikeseries} , with general frequency and the zero-frequency
limit included. The real source pairings, quantitative domains \eqref{eq:spacelikenorm}
and \eqref{eq:spacelikederivativebound} , and explicit FG completion
\eqref{eq:spaceliketensorkernel} specify the extension used here.
Its entire derivative series agrees with the timelike normalizable
branch wherever both representations converge. The covariant series
\eqref{eq:globalFGcontactkernel} gives local FG kernels in conformally
flat representatives under \eqref{eq:globalderivativebound}. Its
global use also requires a regular covering by FG coordinate neighborhoods;
spacelike Fourier support itself remains a flat-frame condition.

Boundary normalizability and convergence do not imply interior regularity.
The present exterior construction consequently admits sources excluded
by the globally regular vacuum spectra. The inclusion of zero frequency
in a controlled, frequency-dependent source family provides a useful
starting point for BMS and celestial comparisons, but the limiting
Bondi data and charges still have to be calculated. Sharper source
bounds could enlarge the sufficient convergence domains used here.
\begin{acknowledgments}
Y.W. and J.Y. thank the Graduate School at Brown University for support.
\end{acknowledgments}

\appendix

\section{Boundary identities and FG improvement\label{app:FGcompletion}}

\subsection{Boundary identities and the endpoint\label{app:Ward}}

All quantities are in the flat frame. The first identity in \eqref{eq:Wardcuts}
follows from the integration by parts used in \citep{Lowe:2024reconstruction},
with $V_{d}=e_{db}w^{b}$ for a constant symmetric tensor $e^{ab}$.
For the identity used to localize the weighted radial integral, instead
use $V_{d}=Qe_{db}w^{b}-w_{d}e_{ab}w^{a}w^{b}$. Its symmetric derivative
is

\begin{equation}
\partial'_{(c}V_{d)}=-Qe_{cd}+\eta_{cd}e_{ab}w^{a}w^{b}.\label{eq:WardV}
\end{equation}

Integrate $\partial'_{c}[\Theta(Q)\widehat{\tau}^{cd}V_{d}]$. Conservation
and tracelessness remove the bulk derivative and trace terms; on $Q=0$,
$V_{d}=-w_{d}e_{ab}w^{a}w^{b}$. This gives the second identity in
\eqref{eq:Wardcuts}. The stated decay assumptions remove the surface
term at infinity.

The lower endpoint need not cancel between the two time sheets. For
a smooth nonnegative radial $\phi$ supported in a spatial annulus,
take

\begin{equation}
\tau_{00}=\Delta\phi,\qquad\tau_{0i}=0,\qquad\tau_{ij}=\delta_{ij}\Delta\phi-\partial_{i}\partial_{j}\phi.\label{eq:endpointexample}
\end{equation}

This stress is conserved and traceless. At the spatial origin its
endpoint is

\begin{equation}
C_{00}=\frac{4}{3}\int\dd^{2}x'\,|\boldsymbol{x}'|^{3}\Delta\phi=12\int\dd^{2}x'\,|\boldsymbol{x}'|\phi>0.\label{eq:noncancel}
\end{equation}

Here $\Delta r^{3}=9r$, and both time sheets contribute with the
same sign. This source lies outside the timelike decay class. For
\eqref{eq:FTtheta}, integrate first over time and angle, then differentiate
the damped Bessel identity $\int^{\infty}_{0}\dd r\,J_{0}(kr)\cos(\omega r)=\operatorname{Re}[k^{2}-(\omega+i0)^{2}]^{-1/2}$.
Applying $-\Box_{p}$ gives the multiplier for $w^{2}\Theta(w^{2})$.
Null-cone contact terms give zero because all derivatives of the allowed
Fourier source vanish there.

\subsection{The boundary-anchored FG improvement\label{app:projection}}

Along the observation normal, the adapted coordinate transformation
has

\begin{equation}
u=z,\qquad x=x_{0},\qquad\partial_{a}x^{c}=E_{a}{}^{c},\qquad\partial_{z}\partial_{a}x^{c}=-zP_{a}{}^{c}.\label{eq:Jacobi}
\end{equation}

The radial components of the pulled-back perturbation vanish on this
line, but the transverse derivatives required in \eqref{eq:FGxi}
are

\begin{equation}
\begin{aligned}\partial_{a}h_{zb} & =-zP_{a}{}^{c}E_{b}{}^{d}H_{cd},\\
\partial_{a}\partial_{b}h_{zz} & =2z^{2}P_{a}{}^{c}P_{b}{}^{d}H_{cd},\\
\partial_{a}\partial_{b}\xi^{z} & =-zP_{a}{}^{c}P_{b}{}^{d}J_{cd}.
\end{aligned}
\label{eq:radialderivatives}
\end{equation}

Since $E$ is a polynomial in $P$, they commute, and

\begin{equation}
\partial_{z}E^{-1}=zE^{-1}PE^{-1},\qquad J'=z^{3}H.\label{eq:Einversederivative}
\end{equation}

After raising the tangent index, the integrand for $\partial_{a}\xi^{b}$
is the radial derivative of $P_{a}{}^{c}(E^{-1}J)_{c}{}^{b}$. Normalizability
sets the lower endpoint to zero. The tangential Lie derivative is
therefore

\begin{equation}
(\Lie_{\xi}G)_{ab}=\frac{2}{z^{2}}E_{(a}{}^{c}P_{b)}{}^{d}J_{cd}.\label{eq:Liepoint}
\end{equation}

Adding the pulled-back perturbation gives \eqref{eq:FGpoint}. Thus
a gauge vector that vanishes on the observation line can still have
a nonzero metric improvement there. Its leading order is $z^{3}$,
preserving $B=\tau$ and the linearized Weyl tensor.

\section{Convergence and covariance of the contact-term kernels\label{app:spacelikeconvergence}}

Substituting $H=\sum_{n\ge0}c_{n}z^{2n+1}(-\Box_{\eta})^{n}\tau$
into the flat spin-two equation gives

\begin{equation}
c_{0}=1,\qquad c_{n}=\frac{c_{n-1}}{2n(2n+3)}=\frac{6(n+1)}{(2n+3)!}.\label{eq:spacelikerecurrence}
\end{equation}

Under \eqref{eq:spacelikederivativebound}, each term satisfies

\begin{equation}
\left|c_{n}z^{2n+1}(-\Box_{\eta})^{n}\tau\right|\le\frac{3Cz}{(2n+1)(2n+3)}\left(\frac{z}{a}\right)^{2n}.\label{eq:spacelikegeometricbound}
\end{equation}

The strict inequality $z<a$ also controls the polynomial factors
from fixed radial derivatives. Applying the bound to $\partial^{\alpha}\tau$
justifies tangential differentiation and the Ward identities term
by term. The sum has $A=0$, $B=\tau$. Radial integration divides
each term by $2n+5$, giving \eqref{eq:spacelikeL}.

The Fourier bound implies the derivative estimate through $q^{2n}\le(2n)!a^{-2n}e^{aq}$.
On $z\le z_{*}<a$, differentiated multipliers are bounded by a polynomial
in $P$ times $e^{z_{*}q}$. These bounds justify the Fourier cutoff
limit, interchange with the contact-term series, and the source limits
in Section \ref{sec:zerofrequency}. The derivative condition also
defines the series on stresses given only in a real neighborhood,
without a global Fourier transform.

To obtain \eqref{eq:covariantD1}, use flat coordinates with $g_{(0)}=e^{2\omega}\eta$
and $\omega=\partial\omega=0$ at the observation point. There $\widehat{\tau}=e^{\omega}\tau$,
$P_{c}{}^{c}=-\Box_{\eta}\omega$, and

\begin{equation}
D^{c}D_{c}\tau_{ab}=\Box_{\eta}\tau_{ab}-2(\Box_{\eta}\omega)\tau_{ab}.\label{eq:covariantD1proof}
\end{equation}

It follows that $-\Box_{\eta}(e^{\omega}\tau)=-D^{c}D_{c}\tau+3P_{c}{}^{c}\tau$.
Expanding $E$ and $L_{z}$ then gives \eqref{eq:globalFGleading}.
Higher orders define the operators $\mathcal{D}_{n}$ by the same
local coordinate and tensor transformations.

The transformed coefficients, including the FG improvement, solve
the curved Einstein equations order by order. Equation \eqref{eq:globalderivativebound}
permits the series and all fixed derivatives to be summed. The two
independent leading behaviors of $z^{2}h$ are $z^{0}$ and $z^{3}$;
fixing $A$ and the conserved, traceless $B$ determines every higher
coefficient. Convergent local expressions with the same data therefore
agree on overlaps. This proves the metric construction for locally
conformally flat representatives within regular FG coordinates. A
global reconstruction requires compatible data and convergent neighborhoods
covering the region of interest; it does not follow from conformal
flatness alone.

\section{Global examples and limitations of light-cone localization\label{app:globalexamples}}

\subsection{A smooth source and its FG improvement\label{app:obstruction}}

Start with a nonzero, even, nonnegative smooth pulse $g$ supported
in $|t|<\delta<\pi/2$, and extend it by $g(t+\pi)=-g(t)$. Put $v=(\partial^{2}_{t}+1)g$
and $Y=P_{2}(\cos\theta)$. The conserved, traceless boundary stress
is

\begin{equation}
\begin{aligned}\tau_{tt} & =vY,\qquad\tau_{ti}=-\frac{\dot{v}}{6}D_{i}Y,\\
\tau_{ij} & =\frac{5v+\ddot{v}}{4}\sigma_{ij}Y+\frac{v+\ddot{v}/3}{4}D_{i}D_{j}Y.
\end{aligned}
\label{eq:bumpstress}
\end{equation}

Antiperiodicity gives odd frequencies, and $\partial^{2}_{t}+1$ removes
$\pm1$, leaving frequencies in the regular scalar $\ell=2$ spectrum.
Summing the corresponding masters of \eqref{eq:regularmasters} gives,
with $\beta=\arctan(1/r)$,

\begin{equation}
\begin{aligned}\Phi(t,r) & =\frac{1}{4}\left[\left(1+\partial^{2}_{t}+\frac{3}{r^{2}}\right)G(t,\beta)+\frac{3}{r}\partial_{\beta}G(t,\beta)\right],\\
G(t,\beta) & =\frac{1}{2}\bigl[g(t+\beta)+g(t-\beta)\bigr].
\end{aligned}
\label{eq:bumpmasterpacket}
\end{equation}

At time zero and $r>0$ with $\beta(r)>\delta$, this master and all
its derivatives vanish near the observation point. The anchored improvement
still integrates outer radii, with $H=3\Phi/r+\Phi_{r}+(r/f)\Phi_{tt}$.
Substitution of $u=\cot s$ and integration by parts give

\begin{equation}
\begin{aligned}\alpha(0,r) & =\frac{\sqrt{f}}{2}\int^{\infty}_{r}\frac{H(0,u)}{\sqrt{f(u)}}\dd u=\sqrt{f}\,A_{g},\\
A_{g} & =\frac{3}{8}\int^{\delta}_{0}\sec^{3}s\,g(s)\dd s>0,\qquad\delta<\beta(r).
\end{aligned}
\label{eq:bumpgaugeintegral}
\end{equation}

Compact support removes the upper integration terms and evenness removes
odd derivatives at zero. On the equator, $Y=-1/2$ and $D_{\phi}D_{\phi}Y=0$,
so

\begin{equation}
h^{{\rm FG}}_{\phi\phi}(0,r,\pi/2)=-r\sqrt{1+r^{2}}\,A_{g}\ne0,\qquad\delta<\beta(r).\label{eq:bumpnonzero}
\end{equation}

On both complete light-cone cuts, $\beta(r)\le|t'|\le\pi-\beta(r)$.
The stress and all its derivatives vanish near those cuts, including
the antipodal images. Every ordinary distribution supported there
gives zero, contradicting \eqref{eq:bumpnonzero}. The surviving FG
component is pure gauge: the master, pre-FG gauge perturbation, and
Weyl curvature vanish locally. This establishes a limitation of the
FG metric on arbitrary smooth sources, not a failure of physical reconstruction.

\subsection{Unmodified gluing for an analytic single mode\label{app:originalgluing}}

Use the same tensor structure \eqref{eq:bumpstress} with $v(t)=\cos3t$.
This is the lowest regular scalar mode, $\ell=2,n=0$, with $\tau_{tt}=Y\cos3t$
and radial master profile $\Phi=r^{3}/[4(1+r^{2})^{3/2}]$. At time
zero on the north polar axis, the true FG response is

\begin{equation}
h^{{\rm FG}}_{tt}=3\sqrt{1+r^{2}}\left[1-r\arctan(1/r)\right]=z+\frac{3}{20}z^{3}+O(z^{5}).\label{eq:fundtrue}
\end{equation}

Unmodified gluing applies the patch weight \eqref{eq:correctedcut}
with the actual Jacobians \eqref{eq:globalchart}-{}-\eqref{eq:globalconvert},
then fills the missing hemisphere by antipodal images without extra
chart-boundary or radial-endpoint terms. A factor $1/2$ prevents
double counting of antipodal pairs. Substituting this source and taking
the boundary limit, with $\mu=\cos\theta'$, gives

\begin{equation}
\begin{aligned}\lim_{z\to0}z^{2}h^{{\rm glued}}_{tt} & =\frac{3}{2}\int^{1}_{-1}\dd\mu\,\sqrt{1-\mu^{2}}\,\mathcal{P}_{1}(\mu)=\frac{3\pi}{2},\\
\mathcal{P}_{1}(\mu) & =-24\mu^{6}+42\mu^{4}-\frac{47}{2}\mu^{2}+\frac{9}{2}.
\end{aligned}
\label{eq:fundwrongA}
\end{equation}

The polynomial integral equals $\pi$; no extra contact term at the
chart boundary is included. This gives a nonzero $A_{tt}$, whereas
\eqref{eq:fundtrue} has $A_{tt}=0$ and $B_{tt}=1$. Thus analyticity
and the antipodal stress relation do not justify that particular gluing
weight. This example disproves the unmodified patch-gluing prescription;
the preceding smooth-source example rules out any ordinary distribution
supported only on the light-cone cuts that reconstructs the FG metric
for every smooth source.

\bibliographystyle{utcaps}
\bibliography{holograv_normalizable}

\providecommand{\href}[2]{#2}\begingroup\raggedright\begin{thebibliography}{10}

\bibitem{Maldacena:1999aa}
J.~Maldacena, ``The Large-N Limit of Superconformal Field Theories and
  Supergravity,'' \href{https://dx.doi.org/10.1023/A:1026654312961}{{\em
  International Journal of Theoretical Physics} {\bfseries 38} no.~4, (1999)
  1113--1133}. \url{https://doi.org/10.1023/A:1026654312961}.

\bibitem{Witten:1998qj}
E.~Witten, ``{Anti-de Sitter space and holography},''
  \href{https://dx.doi.org/10.4310/ATMP.1998.v2.n2.a2}{{\em Adv. Theor. Math.
  Phys.} {\bfseries 2} (1998) 253--291},
  \href{https://arxiv.org/abs/hep-th/9802150}{{\ttfamily
  arXiv:hep-th/9802150}}.

\bibitem{Lowe:2024reconstruction}
D.~A. Lowe, Y.~Wang, and J.~Yang, ``{Holographic Reconstruction of
  Gravitational Perturbations in AdS/CFT and Implications for Celestial
  Conformal Field Theory},''
  \href{https://dx.doi.org/10.1007/JHEP04(2025)070}{{\em JHEP} {\bfseries 04}
  (2025) 070}, \href{https://arxiv.org/abs/2411.02364}{{\ttfamily
  arXiv:2411.02364 [hep-th]}}.

\bibitem{Hamilton:2005ju}
A.~Hamilton, D.~N. Kabat, G.~Lifschytz, and D.~A. Lowe, ``{Local bulk operators
  in AdS/CFT: A Boundary view of horizons and locality},''
  \href{https://dx.doi.org/10.1103/PhysRevD.73.086003}{{\em Phys. Rev. D}
  {\bfseries 73} (2006) 086003},
  \href{https://arxiv.org/abs/hep-th/0506118}{{\ttfamily
  arXiv:hep-th/0506118}}.

\bibitem{Hamilton_2006}
A.~Hamilton, D.~Kabat, G.~Lifschytz, and D.~A. Lowe, ``Holographic
  representation of local bulk operators,''
  \href{https://dx.doi.org/10.1103/physrevd.74.066009}{{\em Physical Review D}
  {\bfseries 74} no.~6, (Sept., 2006) }.
  \url{http://dx.doi.org/10.1103/PhysRevD.74.066009}.

\bibitem{Hamilton:2006fh}
A.~Hamilton, D.~N. Kabat, G.~Lifschytz, and D.~A. Lowe, ``{Local bulk operators
  in AdS/CFT: A Holographic description of the black hole interior},''
  \href{https://dx.doi.org/10.1103/PhysRevD.75.106001}{{\em Phys. Rev. D}
  {\bfseries 75} (2007) 106001},
  \href{https://arxiv.org/abs/hep-th/0612053}{{\ttfamily
  arXiv:hep-th/0612053}}. [Erratum: Phys.Rev.D 75, 129902 (2007)].

\bibitem{Hamilton:2007wj}
A.~Hamilton, D.~N. Kabat, G.~Lifschytz, and D.~A. Lowe, ``{Local bulk operators
  in AdS/CFT and the fate of the BTZ singularity},'' {\em AMS/IP Stud. Adv.
  Math.} {\bfseries 44} (2008) 85--100,
  \href{https://arxiv.org/abs/0710.4334}{{\ttfamily arXiv:0710.4334 [hep-th]}}.

\bibitem{Kabat:2011rz}
D.~Kabat, G.~Lifschytz, and D.~A. Lowe, ``{Constructing local bulk observables
  in interacting AdS/CFT},''
  \href{https://dx.doi.org/10.1103/PhysRevD.83.106009}{{\em Phys. Rev. D}
  {\bfseries 83} (2011) 106009},
  \href{https://arxiv.org/abs/1102.2910}{{\ttfamily arXiv:1102.2910 [hep-th]}}.

\bibitem{kabat}
D.~Kabat, G.~Lifschytz, S.~Roy, and D.~Sarkar, ``Holographic representation of
  bulk fields with spin in AdS/CFT,''
  \href{https://dx.doi.org/10.1103/PhysRevD.86.026004}{{\em Phys. Rev. D}
  {\bfseries 86} (Jul, 2012) 026004}.
  \url{https://link.aps.org/doi/10.1103/PhysRevD.86.026004}.

\bibitem{BhattacharjeeMukherjee:2025}
B.~Bhattacharjee and J.~Mukherjee, ``{Bulk reconstruction for anti-symmetric
  and symmetric gauge fields: $p$-forms and the graviton},''
  \href{https://dx.doi.org/10.1016/j.nuclphysb.2025.116979}{{\em Nucl. Phys. B}
  {\bfseries 1018} (2025) 116979},
  \href{https://arxiv.org/abs/2307.02710}{{\ttfamily arXiv:2307.02710
  [hep-th]}}.

\bibitem{deHaro:2000vlm}
S.~de~Haro, S.~N. Solodukhin, and K.~Skenderis, ``{Holographic reconstruction
  of space-time and renormalization in the AdS / CFT correspondence},''
  \href{https://dx.doi.org/10.1007/s002200100381}{{\em Commun. Math. Phys.}
  {\bfseries 217} (2001) 595--622},
  \href{https://arxiv.org/abs/hep-th/0002230}{{\ttfamily
  arXiv:hep-th/0002230}}.

\bibitem{SerantesWithers:2022}
A.~Serantes and B.~Withers, ``{Convergence of the Fefferman-Graham expansion
  and complex black hole anatomy},''
  \href{https://dx.doi.org/10.1088/1361-6382/aca237}{{\em Class. Quant. Grav.}
  {\bfseries 39} no.~24, (2022) 245010},
  \href{https://arxiv.org/abs/2207.07132}{{\ttfamily arXiv:2207.07132
  [hep-th]}}.

\bibitem{Lowe:2020qan}
D.~A. Lowe and D.~M. Ramirez, ``{BMS Symmetry via AdS/CFT},''
  \href{https://dx.doi.org/10.1007/JHEP01(2021)075}{{\em JHEP} {\bfseries 01}
  (2021) 075}, \href{https://arxiv.org/abs/2007.02851}{{\ttfamily
  arXiv:2007.02851 [hep-th]}}.

\bibitem{deGioiaRaclariu:2024}
L.~P. de~Gioia and A.-M. Raclariu, ``{Celestial sector in CFT: Conformally soft
  symmetries},'' \href{https://dx.doi.org/10.21468/SciPostPhys.17.1.002}{{\em
  SciPost Phys.} {\bfseries 17} (2024) 002},
  \href{https://arxiv.org/abs/2303.10037}{{\ttfamily arXiv:2303.10037
  [hep-th]}}.

\bibitem{BagchiDhivakarDutta:2026}
A.~Bagchi, P.~Dhivakar, and S.~Dutta, ``{3D stress tensor for gravity in 4D
  flat spacetime},'' \href{https://dx.doi.org/10.1103/fmvk-vrdz}{{\em Phys.
  Rev. D} {\bfseries 113} (2026) 066013},
  \href{https://arxiv.org/abs/2408.05494}{{\ttfamily arXiv:2408.05494
  [hep-th]}}.

\bibitem{SkenderisSolodukhin:1999}
K.~Skenderis and S.~N. Solodukhin, ``{Quantum effective action from the AdS/CFT
  correspondence},''
  \href{https://dx.doi.org/10.1016/S0370-2693(99)01467-7}{{\em Phys. Lett. B}
  {\bfseries 472} (2000) 316--322},
  \href{https://arxiv.org/abs/hep-th/9910023}{{\ttfamily
  arXiv:hep-th/9910023}}.

\bibitem{KodamaIshibashi:2003}
H.~Kodama and A.~Ishibashi, ``{A master equation for gravitational
  perturbations of maximally symmetric black holes in higher dimensions},''
  \href{https://dx.doi.org/10.1143/PTP.110.701}{{\em Prog. Theor. Phys.}
  {\bfseries 110} (2003) 701--722},
  \href{https://arxiv.org/abs/hep-th/0305147}{{\ttfamily
  arXiv:hep-th/0305147}}.

\bibitem{DHoker:1999bve}
E.~D'Hoker, D.~Z. Freedman, S.~D. Mathur, A.~Matusis, and L.~Rastelli,
  ``{Graviton and gauge boson propagators in AdS(d+1)},''
  \href{https://dx.doi.org/10.1016/S0550-3213(99)00524-6}{{\em Nucl. Phys. B}
  {\bfseries 562} (1999) 330--352},
  \href{https://arxiv.org/abs/hep-th/9902042}{{\ttfamily
  arXiv:hep-th/9902042}}.

\bibitem{MichalogiorgakisPufu:2007}
G.~Michalogiorgakis and S.~S. Pufu, ``{Low-lying gravitational modes in the
  scalar sector of the global AdS$_4$ black hole},''
  \href{https://dx.doi.org/10.1088/1126-6708/2007/02/023}{{\em JHEP} {\bfseries
  02} (2007) 023}, \href{https://arxiv.org/abs/hep-th/0612065}{{\ttfamily
  arXiv:hep-th/0612065}}.

\bibitem{Bakas:2009}
I.~Bakas, ``{Duality in linearized gravity and holography},''
  \href{https://dx.doi.org/10.1088/0264-9381/26/6/065013}{{\em Class. Quant.
  Grav.} {\bfseries 26} (2009) 065013},
  \href{https://arxiv.org/abs/0812.0152}{{\ttfamily arXiv:0812.0152 [hep-th]}}.

\bibitem{DLMF}
{National Institute of Standards and Technology}, ``{NIST Digital Library of
  Mathematical Functions, Chapters 15 and 18: Hypergeometric Function and
  Orthogonal Polynomials},''
\newblock \url{https://dlmf.nist.gov/}. Accessed 8 October 2026.

\bibitem{ReyRosenhaus:2014}
S.-J. Rey and V.~Rosenhaus, ``{Scanning Tunneling Macroscopy, Black Holes, and
  AdS/CFT Bulk Locality},''
  \href{https://dx.doi.org/10.1007/JHEP07(2014)050}{{\em JHEP} {\bfseries 07}
  (2014) 050}, \href{https://arxiv.org/abs/1403.3943}{{\ttfamily
  arXiv:1403.3943 [hep-th]}}.

\bibitem{HeLysovMitraStrominger:2015}
T.~He, V.~Lysov, P.~Mitra, and A.~Strominger, ``{BMS supertranslations and
  Weinberg's soft graviton theorem},''
  \href{https://dx.doi.org/10.1007/JHEP05(2015)151}{{\em JHEP} {\bfseries 05}
  (2015) 151}, \href{https://arxiv.org/abs/1401.7026}{{\ttfamily
  arXiv:1401.7026 [hep-th]}}.

\bibitem{KapecMitraRaclariuStrominger:2017}
D.~Kapec, P.~Mitra, A.-M. Raclariu, and A.~Strominger, ``{A 2D Stress Tensor
  for 4D Gravity},''
  \href{https://dx.doi.org/10.1103/PhysRevLett.119.121601}{{\em Phys. Rev.
  Lett.} {\bfseries 119} no.~12, (2017) 121601},
  \href{https://arxiv.org/abs/1609.00282}{{\ttfamily arXiv:1609.00282
  [hep-th]}}.

\bibitem{DonnayPuhmStrominger:2019}
L.~Donnay, A.~Puhm, and A.~Strominger, ``{Conformally soft photons and
  gravitons},'' \href{https://dx.doi.org/10.1007/JHEP01(2019)184}{{\em JHEP}
  {\bfseries 01} (2019) 184},
  \href{https://arxiv.org/abs/1810.05219}{{\ttfamily arXiv:1810.05219
  [hep-th]}}.

\bibitem{Pasterski:2021aa}
S.~Pasterski, ``Lectures on celestial amplitudes,''
  \href{https://dx.doi.org/10.1140/epjc/s10052-021-09846-7}{{\em The European
  Physical Journal C} {\bfseries 81} no.~12, (2021) 1062}.
  \url{https://doi.org/10.1140/epjc/s10052-021-09846-7}.

\bibitem{Hijano:2020szl}
E.~Hijano and D.~Neuenfeld, ``{Soft photon theorems from CFT Ward identites in
  the flat limit of AdS/CFT},''
  \href{https://dx.doi.org/10.1007/JHEP11(2020)009}{{\em JHEP} {\bfseries 11}
  (2020) 009}, \href{https://arxiv.org/abs/2005.03667}{{\ttfamily
  arXiv:2005.03667 [hep-th]}}.

\bibitem{AldayNocchiRuzziconiSrikant:2025}
L.~F. Alday, M.~Nocchi, R.~Ruzziconi, and A.~Yelleshpur~Srikant, ``{Carrollian
  amplitudes from holographic correlators},''
  \href{https://dx.doi.org/10.1007/JHEP03(2025)158}{{\em JHEP} {\bfseries 03}
  (2025) 158}, \href{https://arxiv.org/abs/2406.19343}{{\ttfamily
  arXiv:2406.19343 [hep-th]}}.

\end{thebibliography}\endgroup

\end{document}